\documentclass[useAMS,usenatbib]{mn2e}
\usepackage{graphicx}
\usepackage{scalefnt}
\usepackage[]{booktabs}
\usepackage{colortbl}
\usepackage[table]{xcolor}
\usepackage{helvet}
\usepackage{multimedia}

\title[Dynamical evolution of V-type photometric candidates in the outer Main-belt]
{Dynamical evolution of V-type photometric candidates in the outer Main-belt}
\author[M.E. Huaman,V. Carruba, R. C. Domingos]{ M.E. Huaman$^{1}$\thanks{E-mail: mariela@feg.unesp.br}, V. 
Carruba$^{1}$, R. C. Domingos$^{2}$\\
$^{1}$UNESP, Univ. Estadual Paulista, Grupo de din\^{a}mica Orbital e
  Planetologia, Guaratinguet\'{a}, SP, 12516-410, Brazil \\
$^{2}$UNESP, Univ. Estadual Paulista, S\~{a}o Jo\~{a}o da Boa Vista, SP, 
13874-149, Brazil\\
  }

\begin{document}

\date{Accepted ... .  Received ... ; in original form 2014 May 23}

\pagerange{\pageref{firstpage}--\pageref{lastpage}} \pubyear{2010}

\maketitle

\label{firstpage}

\begin{abstract}
V-type asteroids, characterized by two absorption bands at 1.0 and
2.0 $\mu m$, are usually thought to be portions of the crust of
differentiated or partially differentiated bodies.  Most V-type
asteroids are found in the inner main belt and are thought to be current
or past members of the Vesta dynamical family.  Recently, several
V-type photometric candidates have been identified in the central and 
outer main belt.

While the dynamical evolution of V-type photometric candidates 
in the central main belt has been recently investigated, less attention 
has been given to the orbital evolution
of basaltic material in the outer main belt as a whole.  Here we identify
known and new V-type photometric candidates in this region, and study their 
orbital evolution under the effect of gravitational and non-gravitational
forces.  A scenario in which a minimum of three local sources, possibly 
associated with the parent bodies of (349) Dembowska, (221) Eos, and (1459) 
Magnya, could in principle explain the current orbital distribution of V-type 
photometric candidates in the region.
\end{abstract}

%

\section{Introduction}
\label{sec: intro}

Basaltic asteroids are associated with the crust of differentiated 
body. They are usually associated with a V-type spectrum, 
characterized by absorptions bands at 1 and 2 $\mu$ m. 
The majority of basaltic asteroids are found in the inner 
main belt, although a few have also been observed in the outer 
main belt.  The first evidence for another possible source of 
basaltic asteroid in the outer main belt was the asteroid (1459) 
Magnya discovery by Lazarro et al. (2000). Other autors as 
Michtchenko et al. (2002) suggested that Magnya is a fragment 
from a differentiated parent body that broke up in 
the outer belt.

The discovery and analysis of basaltic asteroids independent of Vesta 
can provide insights into the early history of solar system formation 
(Carruba et al. {\bf 2014b, paper I hereafter}).  Since the number 
of originally differentiated
bodies that were scattered into the central and outer main belt 
is not well understood or constrained by previous models, 
and could go from a few (essentially fragments resulting from
cratering events prior
to the formation of the Veneneia crater on Vesta, Milani et al. {\bf 2014}), 
to several dozens (formation and scattering of differentiated
bodies other than Vesta into the central and outer main belt, 
Bottke et al. 2006), setting dynamical constraints
on the minimum number of independent sources of basaltic material
can significantly improve our understanding of this phase of formation
of our Solar System, and is one of the main goals of this paper.

This work, that is the continuation of an analogous study performed in the 
central main belt ({\bf paper I), is  divided as follows}: in Section 2 we 
revise spectral taxonomy of asteroids in the outer main belt and identify 
possible V-type photometric candidates. In Section 3 we define groups of 
V-type candidates and their possible origin. In Section 4 we obtain 
dynamical map of the region of the outer main belt.  Section 5 is dedicated 
to the Yarkovsky orbital evolution, with a subsection on the long-term effect 
of close encounter with massive asteroids.  Finally, we present our 
conclusions in Section 6.

%
\section{Identification: spectral Taxonomy and SDSS-MOC4 data}
\label{sec: taxonomy-sdss}

In this section we use the classification method used in {\bf paper I}, 
and based on the work of DeMeo and Carry (2013).
This method employs SDSS-MOC4 gri slope and ${z}'-{i}'$ colors to classify
an asteroid depending its position on that plane.

As a first part of our analysis, we checked if the preliminary
study of the taxonomy of the candidate V-type asteroids obtained in
{\bf paper I} still holds.  We identified 14 possible V-type candidates
in the outer main belt, 5 of which were spectroscopically observed and
classified as V-types (1459 Magnya, Lazzaro et al. (2000), 7472 Kumakiri, 
Duffard and Roig (2009), 10537 1991 RY16 Moskovitz et al. (2008), 14390 
1990 QP19, De Sanctis (2011a,b), 105041 2000 K041, De Sanctis et al. 
(2011a,b)).  Six objects (11465, 55270, 91159, 92182, 177904, 
and 208324) were reported in Table 1 of {\bf paper I}, while
one (34698 2001 OD22, also listed in the 
WISE albedo catalog\footnote{Initial results from the Wide-field 
Infrared Survey Explorer (WISE) (Wright et al. 2010), and the NEOWISE 
(Mainzer et al. 2011) enhancement 
to the WISE mission recently allowed to obtain diameters and 
geometric albedo values for more than 100,000 Main Belt asteroids 
(Masiero et al. 2011).}) is a newly identified V-type photometric candidate.

V-type asteroid candidate were identified in the past by 
Carvano et al. (2010).  These authors used studies on taxonomy 
of asteroids observed by the Sloan Digital Sky Survey-Moving Object 
Catalog data, fourth (SDSS-MOC4 hereafter), and identified 52 
V-type photometric candidates in the outer main belt, including 
objects with QS, SV, and other cases. This database is available 
on the Planetary Data System (http:sbn.psi.edu/pds/resource/sdsstax.html, 
and was accessed on November $16^{th}$ 2013).

With respect to this database, we selected all asteroids that were 
V-type photometric candidates (including object with QS, SV and 
other cases) and identified according to the HORIZONS System of the 
Jet Propulsion Laboratory (Giorgini et al. 1996).  Except the five confirmed
V-type asteroids, we encountered 25 V-type candidates that are pure V 
(excluding QS and SV cases) and have proper elements as 
reported by the AstDyS site.  Also considering the seven new candidates
identified in this work and in {\bf paper I}, and 
the five confimed V-type objects, 
we have a total of 37 asteroids that are possible or confirmed V-types 
in the outer main belt (with known proper elements).

\begin{figure*}
   \begin{minipage}[c]{0.5\textwidth}
    \centering \includegraphics[width=3.0in]{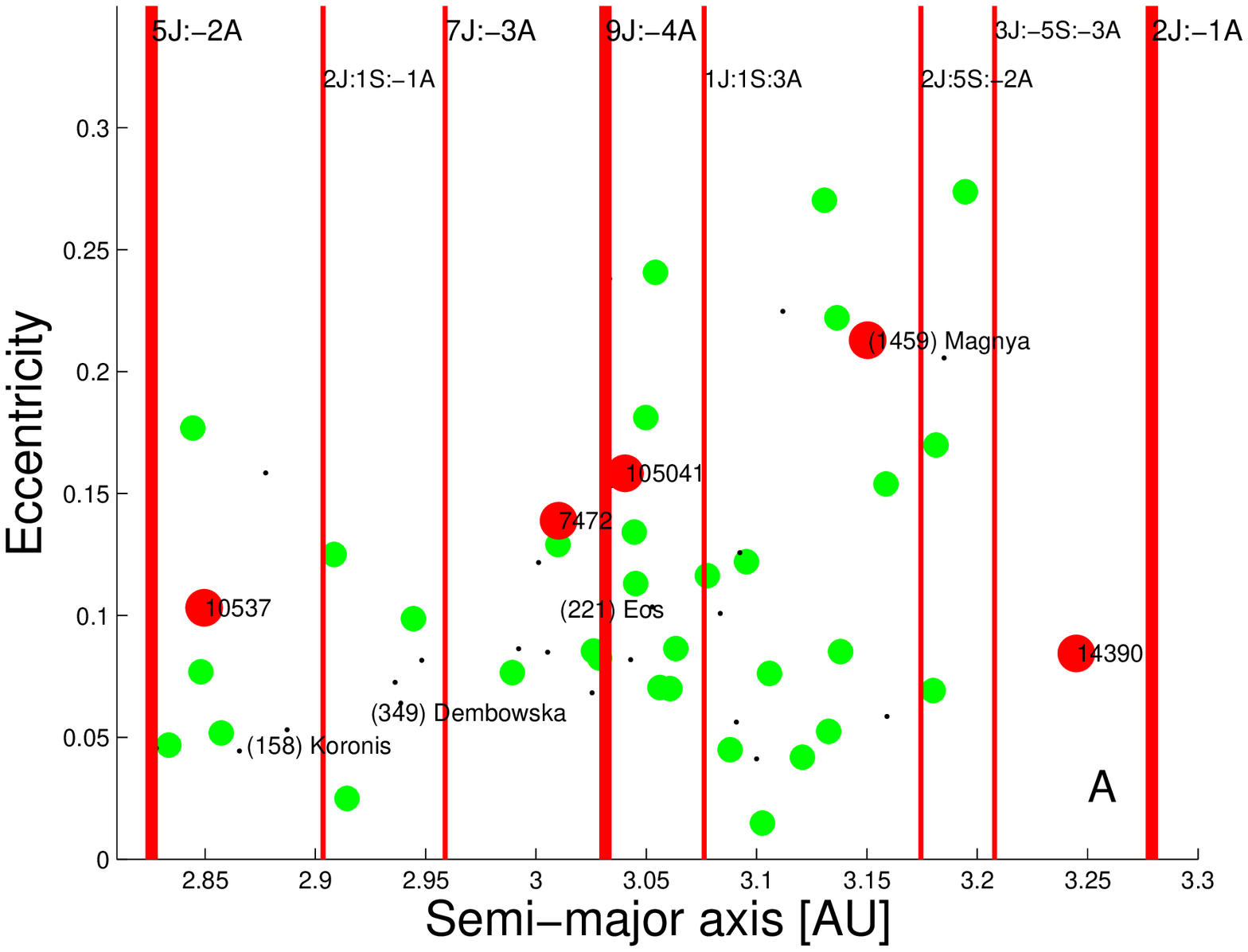}
   \end{minipage}%
   \begin{minipage}[c]{0.5\textwidth}
    \centering \includegraphics[width=3.0in]{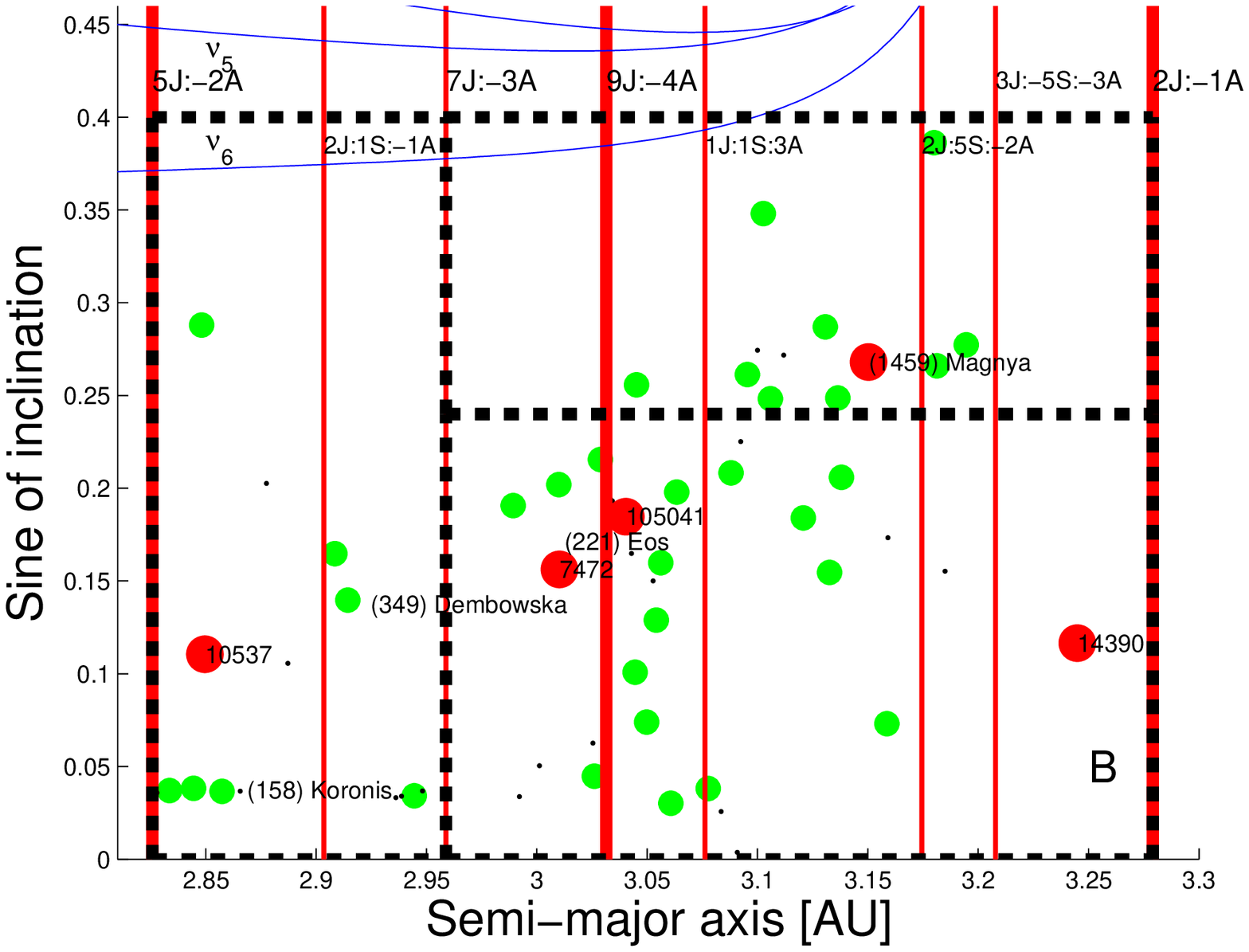}
   \end{minipage}

\caption{Panel A: Proper semi-major axis versus 
eccentricity of the V-type candidates identified using SDSS-MOC4 data 
in the outer main belt. Panel B: Semi-major 
axis versus sine of proper incliation of the V-type candidates.
Vertical lines displays mean-motion resonances, blue lines
show the location of secular resonances, black dotted lines
the limits of the regions in the $(a,sin(i))$ plane.  See text
for a discussion of the other symbols.}
\label{fig: V-type_external}
\end{figure*}

Fig.~\ref{fig: V-type_external} displays a proper $(a,e)$ (panel A),
and $(a,sin(i))$ (panel B) projections of the proper elements of 
all candidates (black dots), the pure V-type candidates 
(green full dots) and of the confirmed V-types asteroids
(red full dots). Vertical lines represent the location of 
the main local mean-motion resonances, blue lines show 
the positions of the center of secular resonances, computed 
using the analytical theory 
of Milani and Kne\v{z}evi\'{c} (1994)
to determine the proper frequencies $g$ and $s$ for the grid of $(a,e)$
and $(a,\sin(i))$ values shown in Fig.~\ref{fig: V-type_external} and the
values of angles and eccentricity of (31) Euphrosyne, the asteroid with the 
largest family in the highly inclined region.

We can observe in Fig.~\ref{fig: V-type_external} 
three regions of concentrations of V-types candidates in 
the $(a, sin(i))$ plane. One region is around Magnya (Magnya region hereafter)
and is between the resonances 7J:-3A and 2J:-1A. This region includes 
the confirmed V-type asteroid (1459) Magnya, with a range values 
of $sin(i)$  of $0.24<sin(i)<0.40$ .
A higher concentration near the Eos family is found in the center of the 
outer main belt (Eos region, since Eos has been suggested
as a possible source of basaltic material in the past, 
Moth\'{e}-Diniz et al. 2008) between the resonances 7J:-3A and 2J:-1A,
with $0.00<sin(i)<0.24$.   Finally, in the region 
between the resonances 5J:-2A and 7J:-3A and $sin(i) < 0.40$,
we found some scattered objects that may have passed through 
the resonance 5J:-2A. Since Moskovitz et al. (2008) suggested
that (349) Dembowska could be a parent body of a differentiated
family, and because of the presence of the Koronis family in the 
region, another possible source of differentiated objects, 
we define this area as the Koronis and Dembowska region.


Four V-type photometric candidates were members of asteroid families,
two in the Koronis and two in the Eos families, respectively.  Following
the approach of {\bf paper I}, we estimated the probability
that such occurrences were produced by a Poisson distribution.
Considering a standard Poisson distribution (see also {\bf paper I},
Eq.~1), the probability that a number of objects be produced by a Poisson 
distribution assuming that the expected number of k occurrence 
in a given interval is given by Eq.~1 in {\bf paper I}.
For our purposes we used  for the expected number of occurrences 
in the given interval
the mean values of objects expected in the regions of 
the Koronis and Eos families, given by Eq.~2 in {\bf paper I}.

We used simple {\bf parallelepipedal} regions for the areas occupied 
by the Koronis and Eos families, defined according to the minimum 
and maximum values in 
$(a,e,sin(i))$ observed for the halos of these two families in Carruba et 
al. ({\bf 2013}), defined as $2.828 < a < 2.987~au, 0.0112 < e < 0.082,  
0.0128 < sin(i)< 0.058$, and $2.915< a < 3.186~au, 0.036 < e < 0.108, 
0.144 < sin(i) < 0.207$, respectively.  $v_{Tot}$, the total volume occupied by 
all photometric candidates, was defined as the region between the 
5J:-2A and 2J:-1A mean motion resonances with Jupiter in semi-major axis
(i.e., $2.8258 < a < 3.279~au$), and eccentricity and $sin(i)$ from zero 
to the maximum value of any V-type candidate in the outer main belt, i.e.,
0.2738 and 0.4.  Using Eq.~1 in {\bf paper I}, we found that the 
probability that the two asteroids in the Koronis family region, and the 
six in the Eos {\bf parallelepipedal} region (four of these objects are not 
members of the Eos dynamical family) could be explained as fluctuations of a 
Poissonian distribution are of $ 4.53\times 10^{−2}$, and $2.53\times 10^{−4}$. 
Considering that the null hypothesis threshold is equal to 
$1.0 \times 10^{−2})$, while the results for the Koronis region may possibly
be compatible with fluctuations of a Poissonian distribution,
those for the Eos region seem more robust and may  
indicate a possible local source.

Since 1-dimensional statistical analysis are somewhat 
dependent on the choice of the regions boundaries, and following
the approach of {\bf paper I}, we also perfomed 
{\bf Mardia's} test (Mardia 1970) on multivariate normality 
for the whole 35 V-type photometric candidates population.
Values of the $A$ and $B$ parameters described in {\bf paper I}
were incompatible with $\chi^{2}$ and a normal distribution, which
excludes that V-type photometric candidates follow a tri-variate 
Gaussian distribution as a whole. In the 
next section we will further analyze the three orbital regions 
defined in this section.


\section{Groups of V-type candidates and their possible origin}
\label{sec: V-type_groups}


\begin{figure*}

   \centering \includegraphics[width=3.0in]{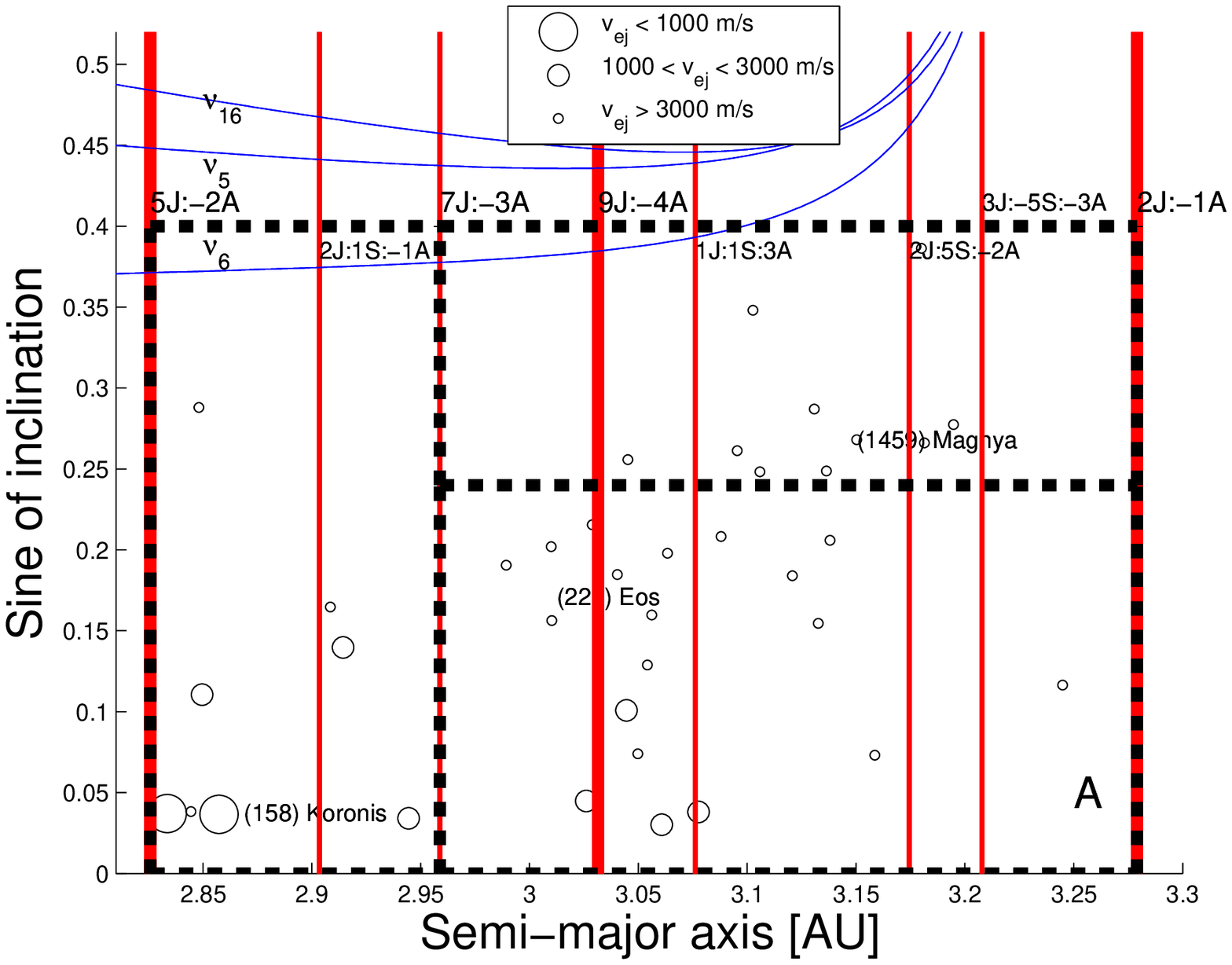}
   \begin{minipage}[c]{0.5\textwidth}
    \centering \includegraphics[width=3.0in]{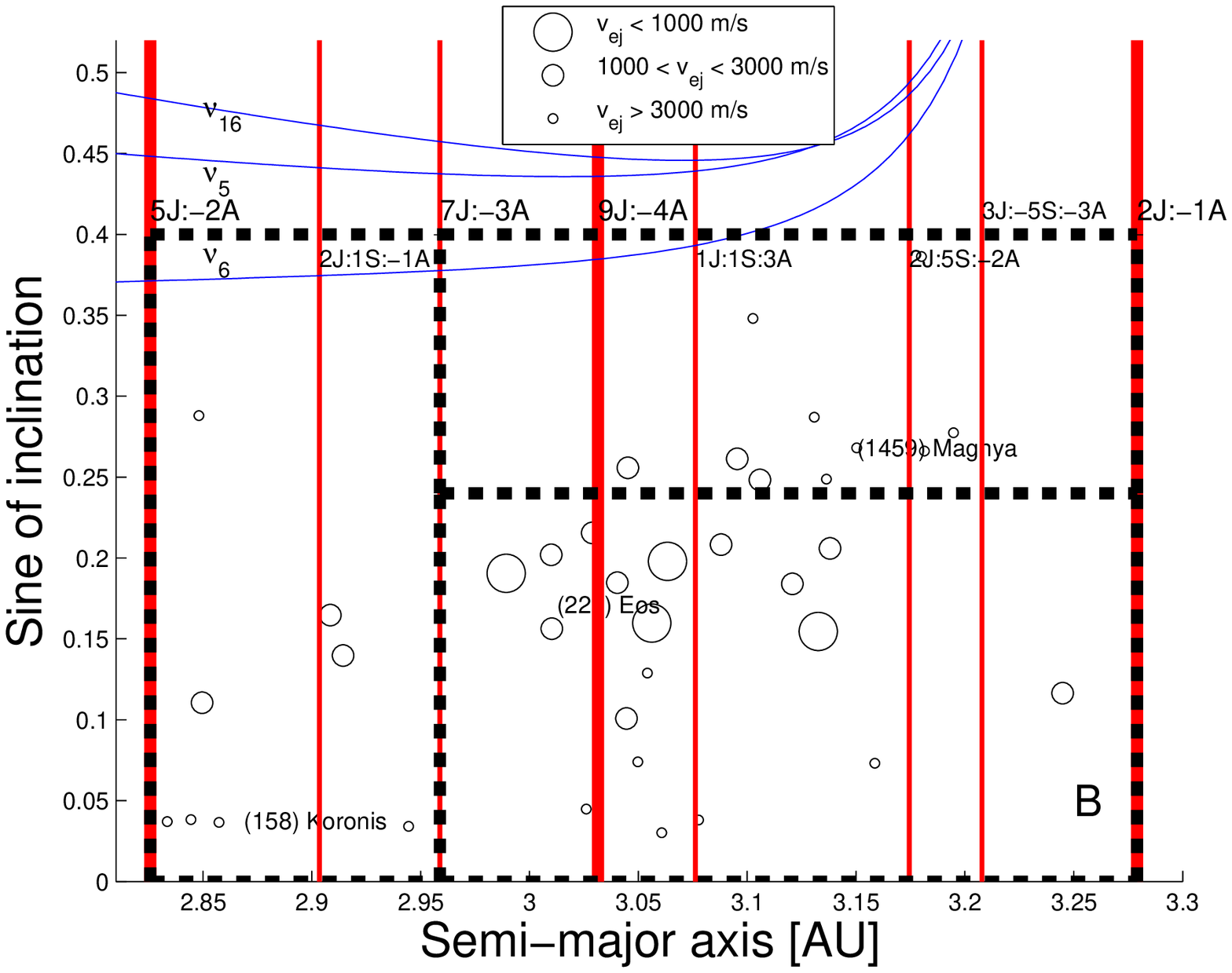}
   \end{minipage}%
   \begin{minipage}[c]{0.5\textwidth}
    \centering \includegraphics[width=3.0in]{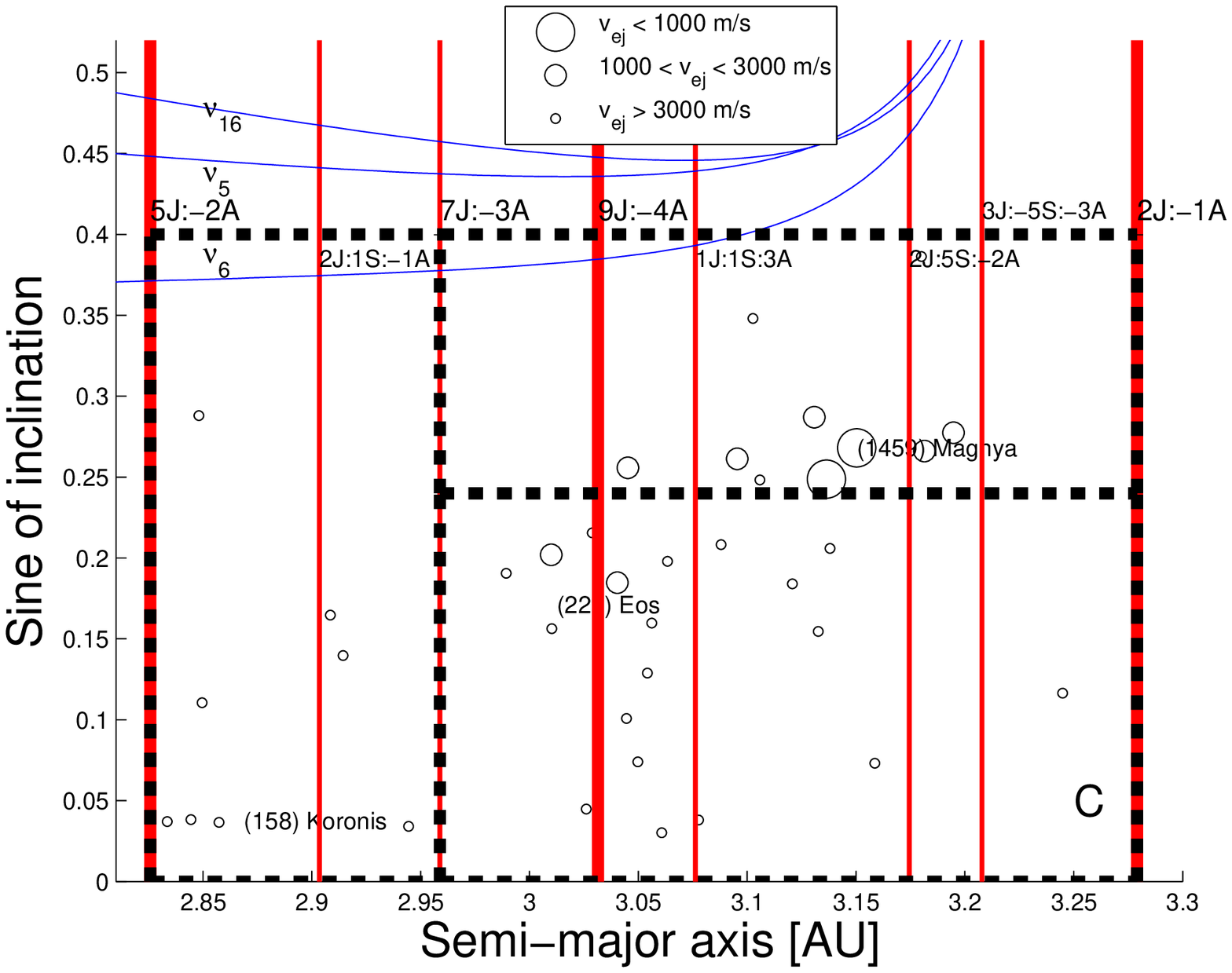}
   \end{minipage}
   
 \caption{An $(a,sin(i))$ projection of V-type photometric candidates
in the outer main belt.  Distances with respect to the asteroids 
(158) Koronis (panel A), (221) Eos (panel B), and (1459) Magnya (panel C)
are shown with the {\bf symbol syze code} discussed in the text.}
\label{fig: velo1} 	
\end{figure*}

To further investigate the robustness of the three regions introduced
in Sec.~\ref{sec: taxonomy-sdss}, we computed the distances
of all photometric V-type candidates in the outer main belt with 
respect to the three suggested possible parent bodies:
(158) Koronis and (349) Dembowska, (221) Eos, and (1459) Magnya.  To 
determine the distances we used the standard
metrics of Zappal\'a et al. (1995) (see also Eq.~7 in {\bf paper I}). 
Results are shown in the three panels of Fig.~\ref{fig: velo1}. 
Asteroids with distances $d$ with respect to (158) Koronis (panel A), 
(221) Eos (panel B), and (1459) Magnya (panel C) less than 
$1000~$m/s are rendered as {\bf large black open circles}, those with 
$1000 < d < 3000~$m/s are shown as {\bf medium-size black circles}, and 
{\bf small circles} identify the objects with $d > 3000~$m/s. 
The other symbols are the same as in Fig.~\ref{fig: V-type_external}. 

Overall, except for a few outliers and two highly inclined
objects (11465 and 34698) in the Magnya region, the closest objects to each 
of the three possible parent bodies tend to be found in its own
region, suggesting that the regions that we proposed in the 
previous section might be a robust feature.

\begin{figure}
  \centering \includegraphics[width=3.0in]{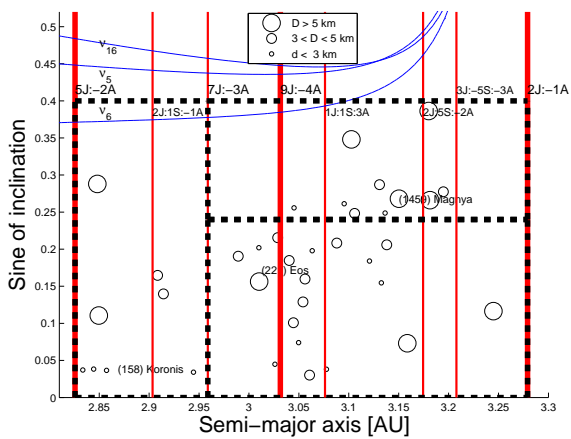}
\caption{An $(a,sin(i))$ projection of V-type photometric candidates
in the outer main belt.   The size {\bf of the symbols} is associated
with the asteroid diameter, according to the figure legend.}  
\label{fig: diameters}
\end{figure}

How much mass is contained in the V-type photometric candidates in the
outer main belt?  To answer this question, we considered the V-type
candidates with diameter values from the WISE mission (12 objects) 
and for the others we computed their diameter using the relationship
of Harris and {\bf Lagerros} (2002) (see also Eq.~8 in {\bf paper I}).  
The minimum and maximum values
of the albedo $p_V$ were 0.040 and 0.392, respectively.  Our results are 
shown in Fig.~\ref{fig: diameters}, where larger asteroids with 
$D > 5$ km are shown as {\bf large black open circle, medium-size 
black circles show the position of the asteroids with $3 < D < 5$~km, 
and the small circles} are associated with smaller bodies. 

To estimate how much mass in the currently known V-type 
candidates is present in the outer main belt, we used the equation 9
from {\bf paper I} (Moskovitz et al. 2008), 
where the bulk density was assumed equal to $3000~$kg/m$^3$, value typical
of V-type objects.  For our candidates, we obtained a 
total mass of $1.75 \times 10^{16}~$kg,  which is 1.35\%
of the estimated mass escavated from craters in Vesta $1.3 \times 10^{18}$~kg; 
more than the 0.139\% found by {\bf paper I} for the
total mass of V-type candidates in the central main belt, but still a 
very small fraction of the observed basaltic material in the inner main belt. 
In the next section we will discuss how V-type photometric candidates 
interact with the local web of mean-motion and secular resonances.


\section{Dynamical Maps}
\label{sec: map}

We continue our analysis by studying the orbital region of 
V-type photometric candidates in the outer main belt. We 
obtained a dynamical map of synthetic proper elements with 18391
test particles using the integrational set-up described in {\bf paper I}. 
We used a step in $a$ of 0.003~{\bf au} and in $i$ of $0.2^{\circ}$ and took
particles in an equally spaced grid of 152 by 121 in the $(a, sin(i))$ plane,
the representative plane for studying diffusion of members
of asteroid families\footnote{Our particles covered
a range between 2.825 and 3.278~{\bf au} in $a$, and $0^{\circ}$ and 
$24.0^{\circ}$ in $i$, respectively.}.  The initial 
values of $e, \Omega, \omega,$ and $\lambda$ of the test particles
were fixed at those of (221) Eos, the largest member of its family
and a possible source of V-type asteroids.  
We then computed synthetic proper elements $(a, e, sin(i))$ 
and frequencies $(n,g,s)$ of these test particles with the approach 
described in Carruba (2010).

\begin{figure*}

  \centering
  \centering \includegraphics [width=0.95\textwidth]{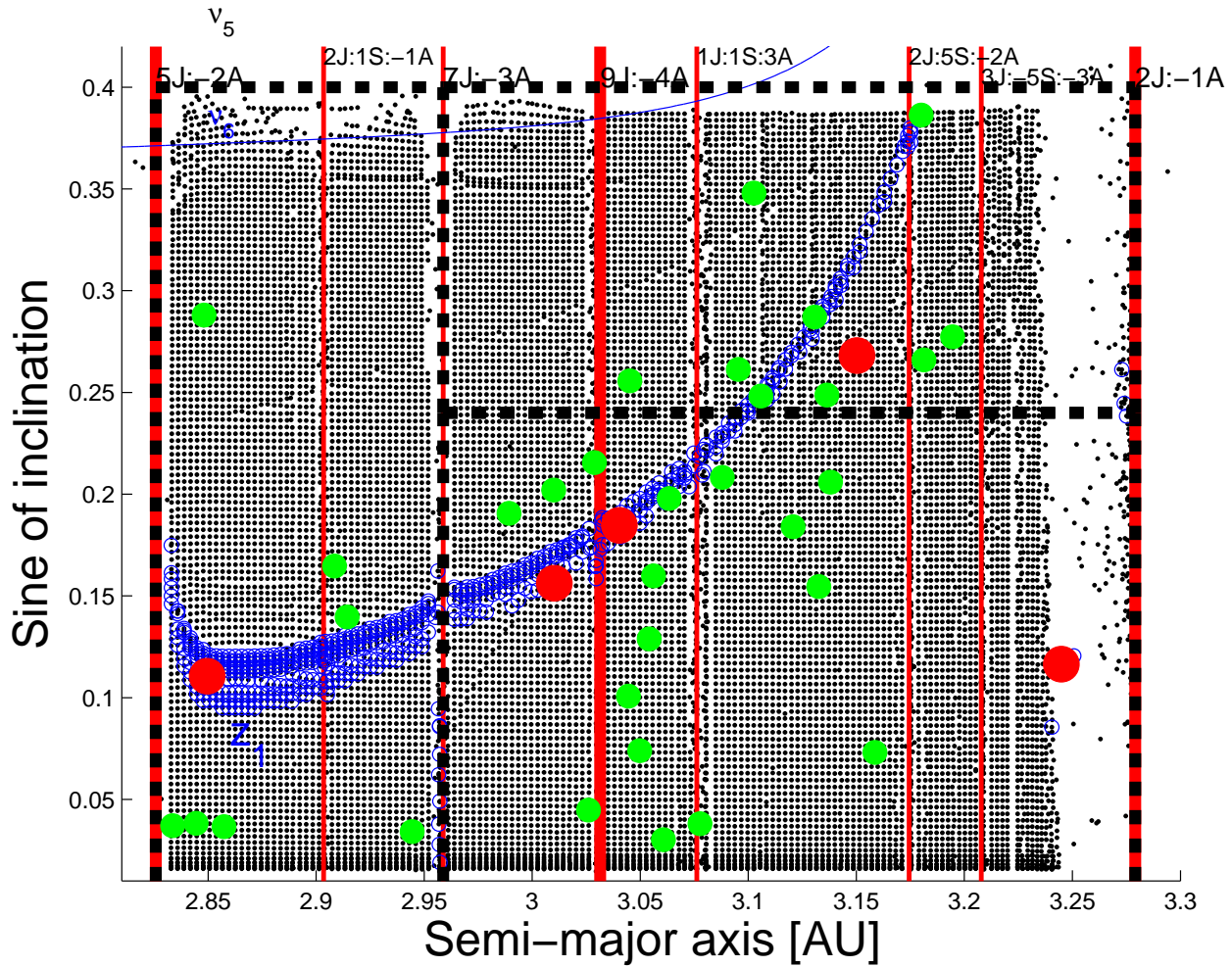}

\caption{An $(a,sin(i))$ proper element map of the outer main belt.
Black dots identify the orbital position in the plane of proper 
$(a, sin(i))$ of each simulated test particle.  {\bf Blue open} circles are 
associated with asteroids likely to be in $z_1$ secular resonance 
configurations. The other symbols have the same meaning as in 
Fig.~\ref{fig:  V-type_external}.} 
\label{fig: map_ai_outer}
\end{figure*}

Fig.~\ref{fig: map_ai_outer} displays a projection in the $(a,sin(i))$
plane of the results of our simulations.  Black dots are associated
with the orbital position of each simulated particle.  Mean-motion resonances
appear as vertical bands with lesser density of test particles~\footnote{Of 
particular interest is the confirmed V-type asteroid
14390  (1990 QP19), that is on the edge of the stable region near the 
2J:-1A mean-motion resonance separatrix.}, secular
resonances show as inclined alignments of test particles.  Other symbols
in the figure have the same meaning as in Fig.~\ref{fig:  V-type_external},
panel B.  The main secular resonance in the outer main belt to appear in 
our map is the $z_1 = g-g_6 +s-s_6$ secular resonance.  We selected
test particles whose combination of asteroidal secular frequencies 
are to within 0.5 arcsec/yr with respect to the resonance
center (i.e., g+s = $g_6+s_6$): these particles are identified by blue
circles in Fig.~\ref{fig:  V-type_external}.  Other secular resonances
are present in the outer main belt but plays a lesser role
when the Yarkovsky force is considered:  except for the $z_1$ and
$2{\nu}_5-2{\nu}_6+{\nu}_{16}$ resonances, Carruba et al. (2014a) showed
that the fraction of objects in librating states of other resonances
drops to less than 40\% of the original population after 100 Myr.

Nevertheless, we studied the time dependence of the resonant argument of 
resonances that showed to be dynamically important in Carruba et al. (2014a)
for the whole 37 V-type photometric candidates populations that we 
identified in Sect.~\ref{sec: taxonomy-sdss}.  In particular, we observed
the time-dependence of the resonant angles of the resonances $z_1, z_2, z_3, 
2{\nu}_5-2{\nu}_6+{\nu}_{16}, 3{\nu}_6-2{\nu}_5, 2{\nu}_6-{\nu}_5$, and
${\nu}_5+{\nu}_{16}$.  With respect to the central main belt, where
{\bf paper I} identified a population of 28 V-type photometric
candidates in secular resonant states, 15 of which inside the 
$2{\nu}_6-{\nu}_{5}$ resonance, in the outer main belt we only
found one object (85812 1998 WR22) inside the $3{\nu}_6-2{\nu}_5$ 
resonance.  Three objects were in circulating orbits  
close to the separatrix of the $z_1$ resonances, and three other 
asteroids were also near the separatrix of the $2{\nu}_5-2{\nu}_6+{\nu}_{16}$
resonance.  While past interactions with secular resonances may of 
course not be excluded, overall secular dynamics seems to have 
played a lesser role in the dynamical evolution of V-type 
photometric candidates in the central main belt with respect 
to the outer belt.  Further assessing the truth of this statement 
will be one of the subjects of the next section.

\section{Dynamical evolution}
\label{sec: yarko}

To investigate the dynamical evolution of the 37 V-type candidates in the 
outer main belt, we used the same methods and procedures presented
in ({\bf paper I}. We integrated clones of these objects with 
SWIFT-RMVSY, the symplectic integrator of Bro\v{z} (1999). 
As in {\bf paper I}, we used the two sets 
of spin axis orientations with $\pm 90^\circ$ with respect to
the orbital plane that maximize the velocity of the drift caused by 
the Yarkovsky effect, and did not considered re-orientations.
To further investigate the dynamical evolution of V-type candidates in 
the outer main belt, we performed 'fast'  and 'slow' numerical simulations 
of these asteroids, assuming that all asteroids had 100 m diameters,
and then using the real diameters. Our results are presented in 
Figs.~\ref{fig: yarko} and \ref{fig: yarko_2},  which
refer to the final orbital evolution on the  $(a-sin(i))$ plane of 
V-type photometric candidates in the region of Eos, Koronis and Magnya.

\subsection{Results of the “fast simulations”}
\label{sec: yarko_fast}

\begin{figure}
   \begin{minipage}[c]{0.5\textwidth}
    \centering \includegraphics[width=3.0in]{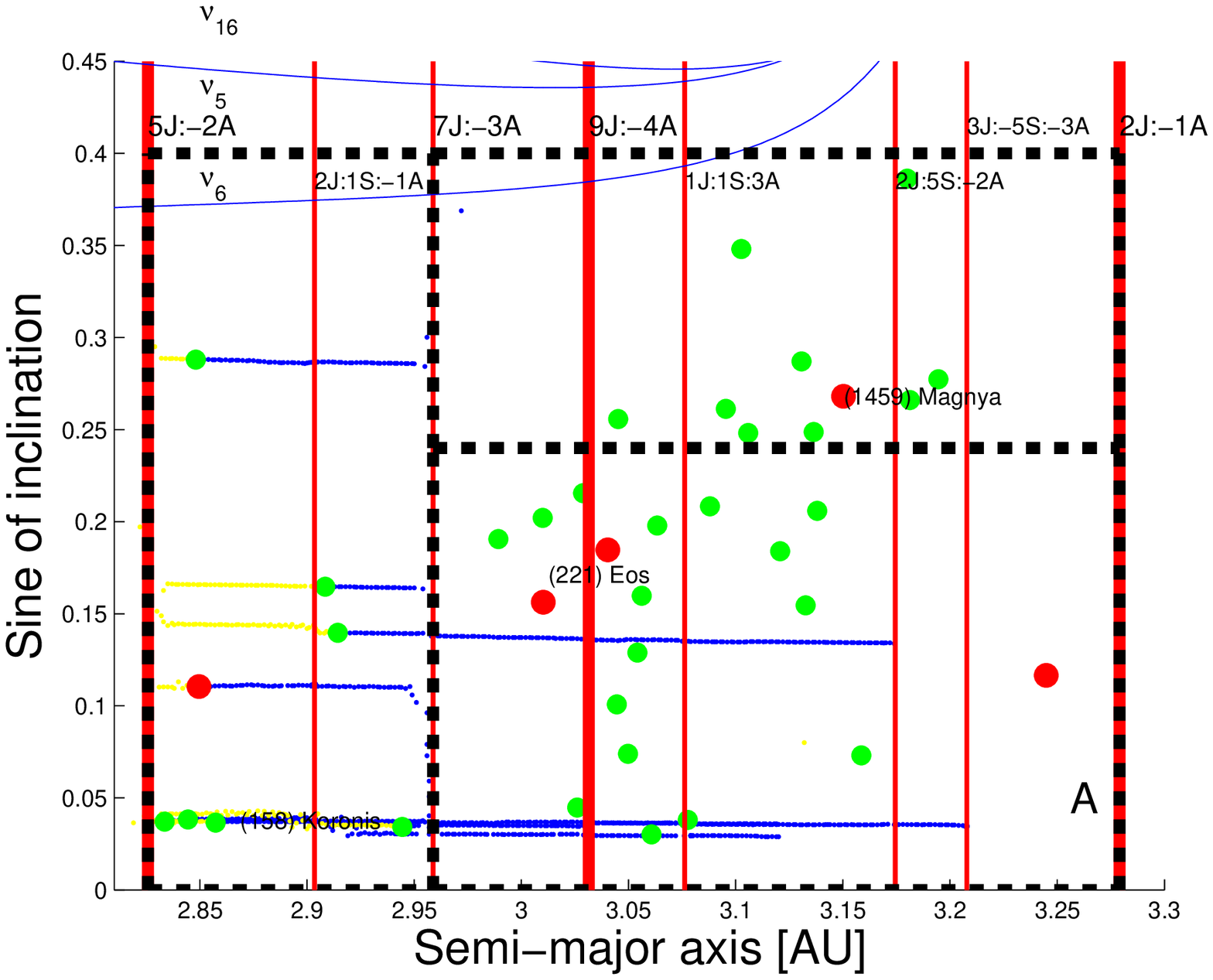}
   \end{minipage}
   
   \begin{minipage}[c]{0.5\textwidth}
    \centering \includegraphics[width=3.0in]{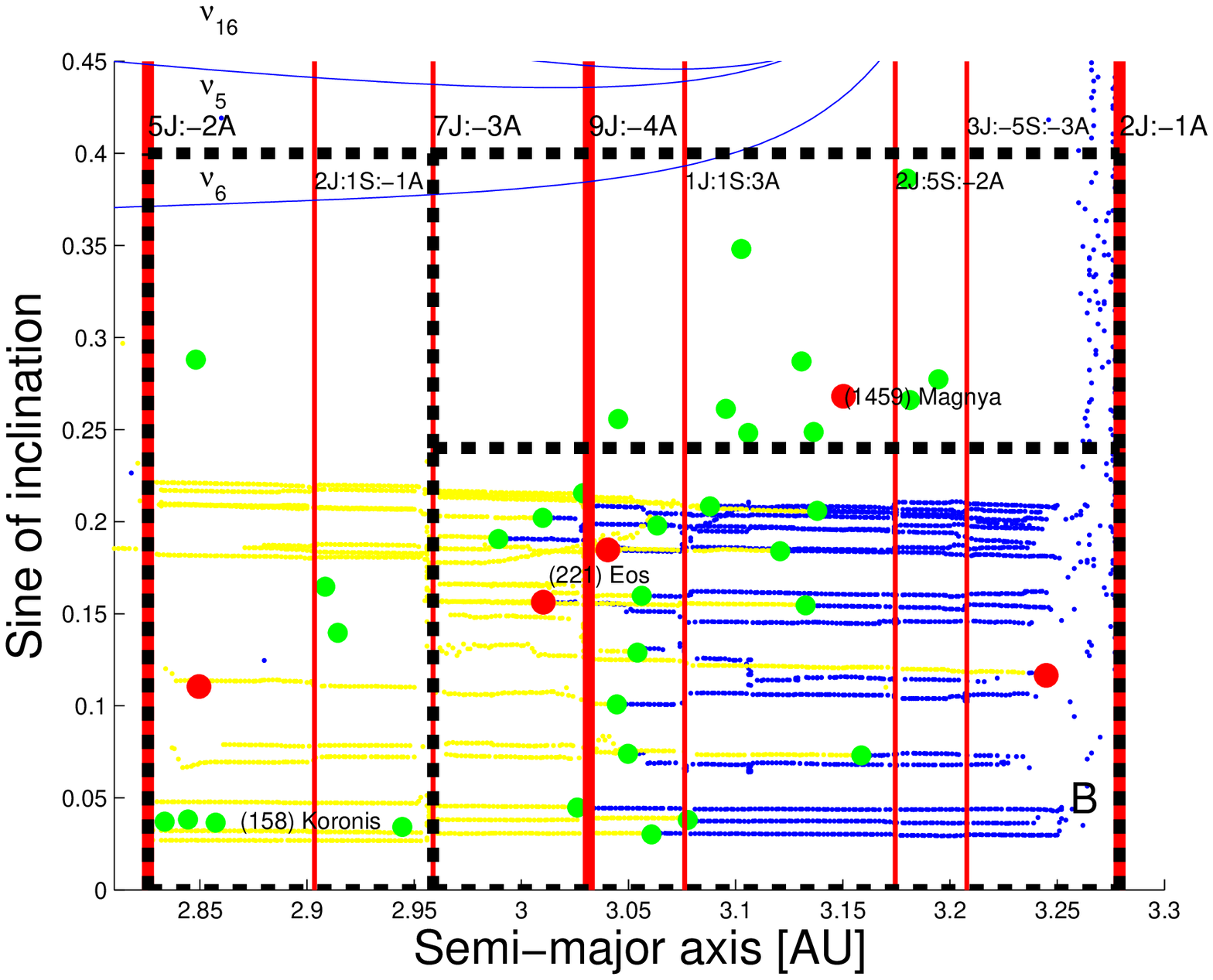}
   \end{minipage}%
   
   \begin{minipage}[c]{0.5\textwidth}
    \centering \includegraphics[width=3.0in]{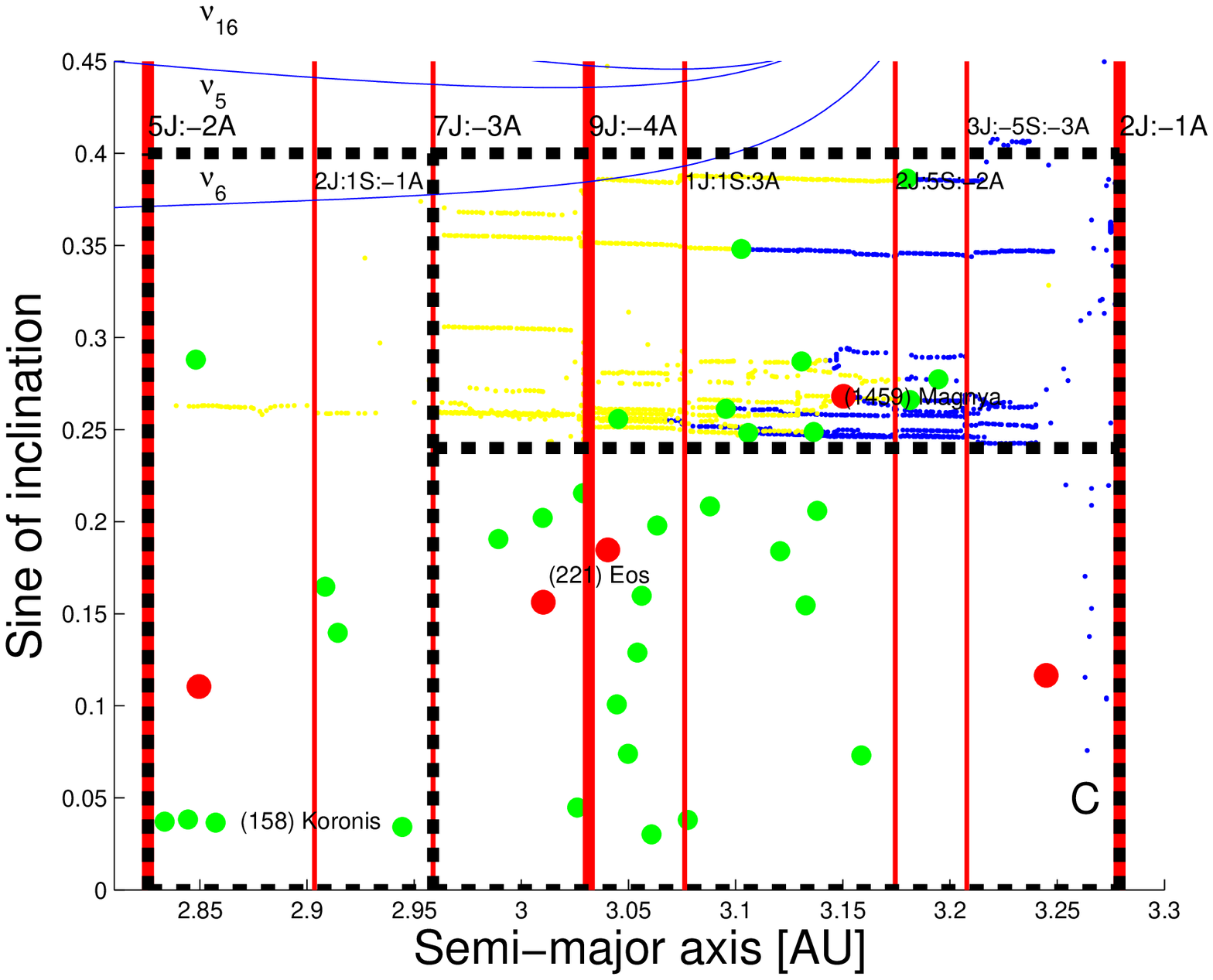}
   \end{minipage}
\caption{Orbital evolution of 100 m clones of V-type photometric 
candidates in the region of Eos, Koronis and Magnya.  See text for further
details on the symbols used in this figure.}
\label{fig: yarko}
\end{figure}

We obtained synthetic proper elements every 1.2 Myr for all 
the simulated asteroids, with the approach 
used in {\bf paper I}, over 200 Myr.  In 
Fig.~\ref{fig: yarko} we show the dynamical paths of clones of 
particles in the Dembowska region (panel A), Eos (panel B) and 
Magnya (panel C) in the $(a,sin(i))$ plane.  
Blue dots represent snaphsots of the orbital 
evolution of clones of real asteroids with $0◦^\circ$ obliquity, while 
yellow dots are associated with clones with $180◦^\circ$ obliquity.  

In the Koronis region there were eight photometric candidates.  
Four of them were not able to cross 
the 7J:-3A mean-motion  resonance, including the (10537) V-type confirmed
body.  This could suggest that communication between the Koronis and
Eos region is in principle possible, even if on long time-scales.
In panel B we observe the orbital evolution of 19 asteroids  in the Eos
region: only 3 of them were not able to cross the 7J:-3A mean-motion 
resonance,  and all clones were able to cross the 
9J:-4A and  3J:-5S:-3A mean-motion resonances, 
until reaching the powerful 2J:-1A mean-motion resonance.
Panel C deals with the orbital evolution of the 10 photometric candidates
in the Magnya region.  Three objects were not able to cross the 
9J:-4A mean-motion resonance, and only one managed to pass the 7J:-3A
resonance.  No asteroids crossed the 2J:-1A mean-motion resonance
and the ${\nu}_6$ secular resonance.  Overall, we found one particle 
that managed to cross the main dynamical 
barrier of the 5J:-2A resonance and none that crossed the  
2J:-1A mean motion resonance, suggesting that comunication 
between the central and outer main belt might be possible, but not likely 
(how effective is this mechanism for larger asteroids will also be
discussed in Sect~\ref{sec: yarko_long}.


Finally, we also observed minor changes of inclination caused by 
secular resonances (mostly by the $z_1$ resonance) 
for some particles, but not enough to allow them to change region.
Based on the results of these ``fast simulations'' we may conclude
that communication between the Koronis and Eos regions and between
the Koronis and Magnya regions may be possible, but that passing 
from the Eos to the Magnya region and vice-versa might a more unlikely 
event.  We will further investigate this hypothesis in the next subsection. 

\subsection{ Results of the “slow simulations”}
\label{sec: yarko_long}

To further investigate the dynamical evolution of V-type candidates 
in the outer main belt, we also performed “slow” simulations of clones of 
the same particles studied in Sect~\ref{sec: yarko_fast}. We took 
the WISE values of the diameters for the test particles for 
which such information was available, and we used the methods 
described in Sect~\ref{sec: V-type_groups} for the other particles.  
We also took six values of spin obliquities, $0^\circ$ , $30^\circ$ , 
$60^\circ$ , $120^\circ$ , $150^\circ$ , and $180^\circ$, in order to 
sample different speeds of drifts caused by the Yarkovsky force.  
We integrated our test particles over 1 Byr, with the 
same integration scheme used in Sect~\ref{sec: yarko_fast}. 
As in {\bf paper I}, because of the longer integration 
time used in these 
runs, we computed proper elements every 4.9 Myr instead of 
the 1.2 Myr used in Sect~\ref{sec: yarko_fast}.


\begin{figure}
   \begin{minipage}[c]{0.5\textwidth}
    \centering \includegraphics[width=3.0in]{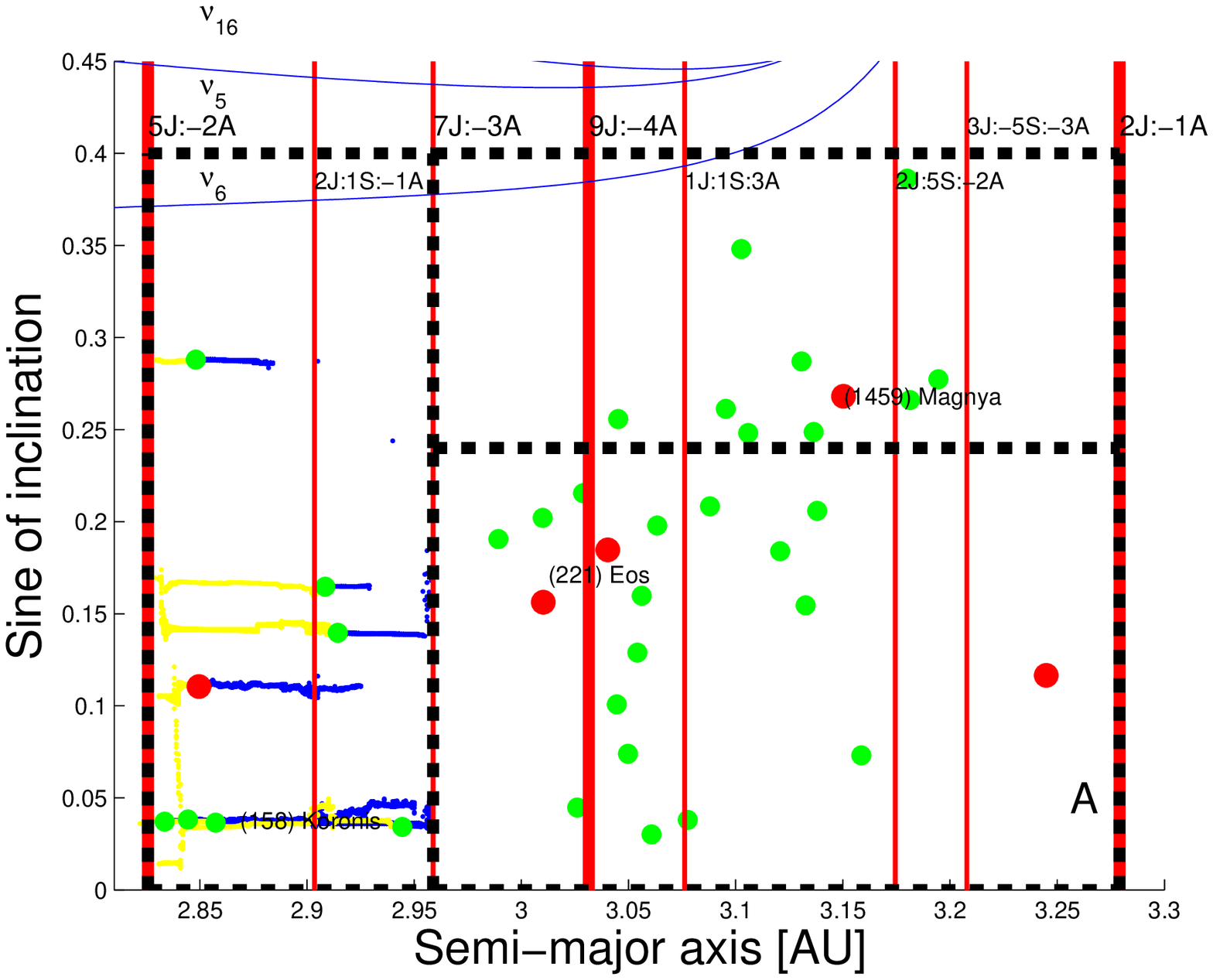}
   \end{minipage}%

   \begin{minipage}[c]{0.5\textwidth}
    \centering \includegraphics[width=3.0in]{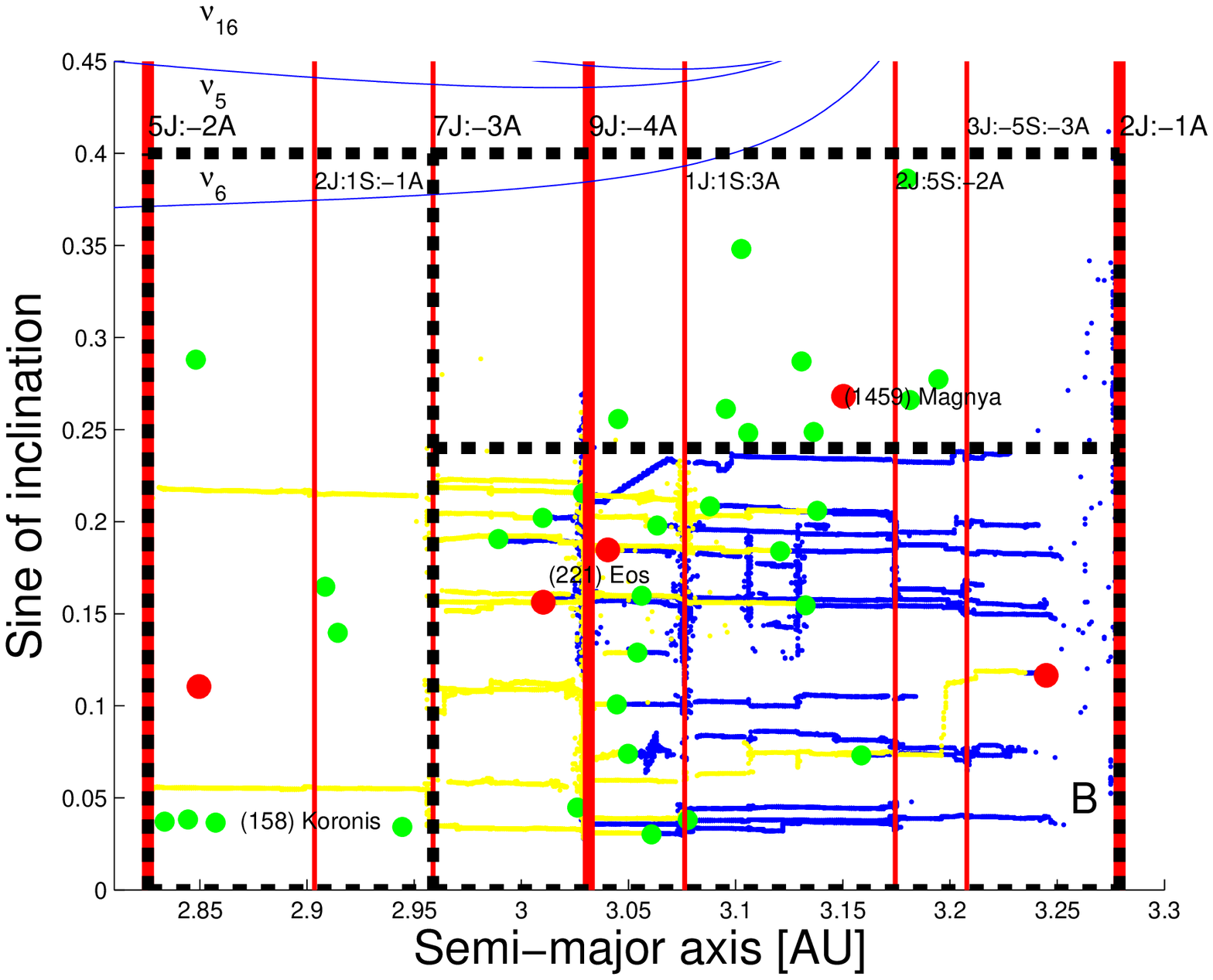}
   \end{minipage}%
   
   \begin{minipage}[c]{0.5\textwidth}
    \centering \includegraphics[width=3.0in]{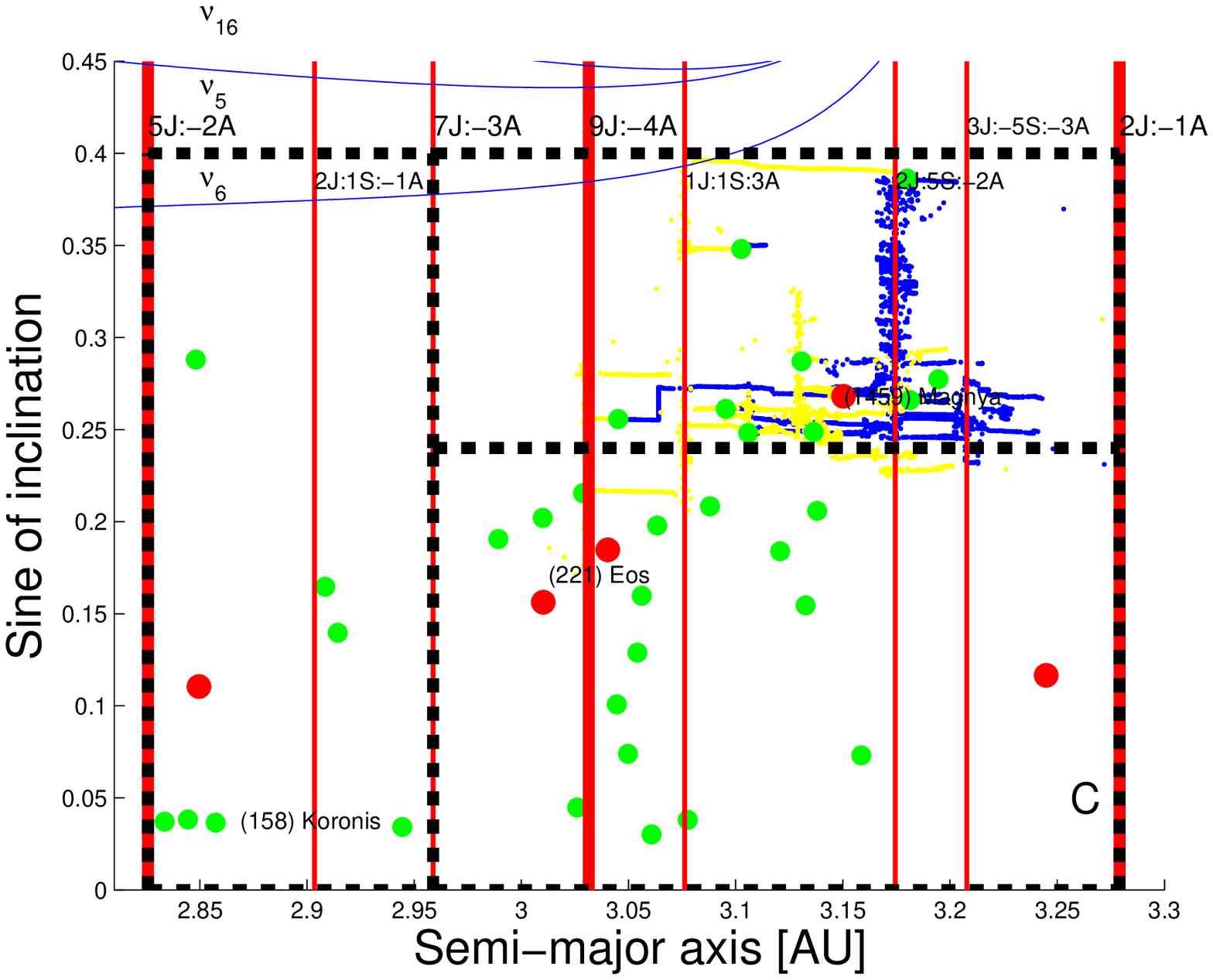}
   \end{minipage}
\caption{Orbital evolution of clones of real asteroids in the region 
of Koronis, Eos, and Magnya, integrated over 1 Byr.  
Other symbols are the same as those used in Fig.~\ref{fig: yarko}.}
\label{fig: yarko_2}
\end{figure}


Fig.~\ref{fig: yarko_2} displays the time evolution of proper
$(a,sin(i))$ values for clones of V-type
photometric candidates with obliquities of $60^{\circ}$ and $150^{\circ}$
in the Koronis region, with {\bf obliquities} of $0^{\circ}$ and $180^{\circ}$
in the Eos region, and  with obliquities of $60^{\circ}$ and $150^{\circ}$ in 
the Magnya region (panels A, B, and C, respectively).   Prograde 
{\bf rotation} clones elements are shown as blue dots, 
retrograde {\bf rotation} clones ones are identified as yellow dots.  
Other symbols are the same as those used in Fig.~\ref{fig: yarko}.  Results 
are similar for other values of the obliquities, and will not be shown for 
the sake of brevity.

Of the 222 integrated clones, none was able to go from the Koronis 
region to the Eos and Magnya areas or to cross the 7J:-3A mean-motion 
resonance.  Only 2 particles from the Eos region managed to cross the 
7J:-3A mean-motion resonance and go into the Koronis area, i.e., 1.8\% 
of the total, suggesting that communication 
across the 7J:-3A mean-motion resonance should be a relatively rare event.

More difficult to interpret were the results near the boundaries of the 
Magnya and Eos region.   Of the 114 integrated clones in the Eos area, 
3 particles (2.7\%) managed to reach the Magnya area and survive 
(mostly because of interaction with the $z_1$ secular resonance). 
Conversely, of the 60 particles integrated in the Magnya region, 
only 5 (4.5\%) were able to reach the Eos region via diffusion in 
the $z_1$ secular resonance and interaction with other mean-motion 
resonances in the area, and remain there stably.
In all the six simulations, we only observed 3 particles that 
managed to cross the 9J:-4A main-motion resonance and approach the 7J:-3A,
resonance, but were not able to cross it.  (1459) Magnya is confined in a zone 
defined by the $z_1$ and 2J:-1A resonances, and no particles managed 
to cross the ${\nu}_6$ {\bf secular} resonance.

Overall, communication between the Eos and Magnya region seems possible,
but not on scale large enough to dismiss the two source scenarios for these
regions.  While some mixing among V-type photometric candidates coming 
from the Magnya and Eos parent body could be possible, our results still 
seems to suggest a minimum of three-sources to explain the current population
of V-type photometric candidates in the outer main belt. 

\subsection{Effect of close encounter with massive asteroids}
\label{sec: encounter}

{\bf Paper I} recently estimated the long term effect of 
diffusion caused by close encounters with three massive asteroids
in the central main belt, including (1) Ceres, the most massive
body in the main belt, and by far the most effective perturber.
Under the assumption that long-term effect of close encounters
can be modeled, in first approximation, as a random walk with
zero mean, there will be a $1\sigma$ (= 68.27\%) probability that  
the root mean square translation distance (or quadratic mean) after $n$ 
steps will fall between $\pm \sigma_{sin(i)} \sqrt{n}$, where $\sigma_{sin(i)}$
is the standard deviation of changes in $\sin(i)$, equal to 0.0029 for
the distribution of changes in $i$ caused by (1) Ceres over 200 Myr.
For 4 Byr, the maximum time for which Bottke et al. (2006) estimated
the possible arrival of differentiated bodies in the main belt, 
the root mean square translation in $\sin(i)$ would be of 
0.0130. Considering our boundary between the inclination regions of Eos 
and Magnya at $\sin(i) = 0.24$, there are only 2 Magnya objects 
((63075) 2000 WC126 and (188331) 2003 OZ32) that could have changed
inclination region over 4 Byr only because of encounters with massive
asteroids, i.e., 5.4\% of the total.  Overall, close encounters
with massive asteroids may have furnished an additional mechanism 
for dynamical mobility, but alone could not have produced substantial
mixing between the Eos and Magnya regions.
  
\section{Conclusions}
\label{sec: concl}

The main goal of this article was to study possible diffusion paths
of V-type photometric candidates in the the outer main belt. 
For these purposes, we: 

\begin{itemize}

\item Revised the current knowledge on confirmed 
and pure V-type photometric candidates from SDSS-MOC4 data in the area.
Using the DeMeo and Carry (2013) approach, we identified 7 new V-type
photometric candidates in the region for a total of 37 possible V-type 
asteroids in the outer main belt.  Six asteroids appear to be members
of the Eos family, and statistical considerations suggest that Eos could be 
a possible local source of basaltic material.

\item Identified three regions in the $(a,sin(i))$ plane associated
with three possible local sources of basaltic material: the parent bodies
of (349) Dembowska, (221) Eos, and (1459) Magnya.   We computed distances
of the V-type photometric candidates and masses of these objects.   The 
closest objects to each of the three possible parent bodies tend to be found
in its own regions, suggesting that these regions may be robust features.
Only 1.35\% of the mass escavated from the craters in Vesta is currently 
present in the outer main belt.

\item Obtained a synthetic proper element map for 18391 particles in the outer
main belt.  The $z_1$ secular resonance is the main non-linear secular resonance
in the area, and we found only one asteroid (85812 WR22) in non-linear
secular librating states, inside the $3{\nu}_6-2{\nu}_5$.

\item Studied the dynamical evolution of clones of V-type photometric 
candidates in the area under the influence of Yarkovsky and YORP forces, 
and estimated the long-term effect of close encounters with massive asteroids.
Overall, our results seems to indicate a possible three sources
scenario for the origin of basaltic material in the outer main belt,
possibly associated with the parent bodies of (349) Dembowska, (221) 
Eos, and (1459) Magnya.  While communication across the 7J:-3A mean-motion 
resonance seems to be a rare event, up to $\simeq 20\%$ of bodies from the 
Magnya region could have reached the Eos area.  Mixing between the Magnya 
and Eos population could be possible, but not on scales large enough to
rule out a three source scenario.  

\end{itemize}

Overall, the main result of our analysis suggest that a minimum three sources
scenario is needed to explain the current orbital distribution of V-type
photometric candidates in the outer main belt.  Contrary to the case of the 
central main belt, mixing between different zones material may occur in 
somewhat larger proportion (up to 5\%), but communication across the 
7J:-3A mean-motion resonance seems an unlikely event for km-sized asteroid. 
A study of mineralogical properties of V-type photometric candidates in 
the region could in principle provide further clues to the possible 
origin of basaltic material in the outer main belt, and on our 
three-sources scenario hypothesis.


\section*{Acknowledgments}

We thank an anonymous reviewer for comments and suggestions 
that significantly improved the quality of this paper.
We would like to thank the S\~{a}o Paulo State Science Foundation 
(FAPESP) that supported this work via the grant 11/19863-3, and the
Brazilian National Research Council (CNPq, grant 305453/2011-4).
This publication makes use of data products from the Wide-field 
Infrared Survey Explorer, which is a joint project of the University 
of California, Los Angeles, and the Jet Propulsion Laboratory/California 
Institute of Technology, funded by the National Aeronautics and Space 
Administration.  This publication also makes use of data products 
from NEOWISE, which is a project of the Jet Propulsion 
Laboratory/California Institute of Technology, funded by the Planetary 
Science Division of the National Aeronautics and Space Administration.

\section{Appendix}
\label{sec: app}

Table~\ref{table: aster_V2} reports the asteroid identification, 
its proper $a,e,$ and $sin(i)$, its absolute magnitude ($H$), 
diameter ($D$), and geometric albedo ($p_V$), according to the WISE mission, 
when available, for all asteroids in the central main belt identified in Carvano
et al. (2010).   Asteroids without WISE data on $D$ and $p_V$ have been
assigned the mean value of geometric albedo of V-type
asteroids in the outer main belt, and diameters computed according 
to Eq. 8 in {\bf paper I} (for simplicity, 
we do not report these data in the Table). 

\begin{table*}
\begin{center}
\caption{{\bf List of V-type photometric candidates in the 
outer main-belt.}}
\label{table: aster_V2}
\vspace{0.02cm}
\begin{tabular}{r c c c c c c c}        
\hline\toprule 
\# Asteroid id. & $a$ & $e$ & $sin(i)$  & H & D(km)& $P_{v}$& Family\\
\midrule 

 \textbf{(349) Dembowska - (158) Koronis}\\
\hline 
   10537&   2.8496&     0.1032&     0.1105&    12.41&     7.8650&   0.3131&\\
   36590&   2.8336&     0.0468&     0.0371&    14.64&     2.7020&   0.1674& Koronis\\
   55270&   2.9085&     0.1250&     0.1647 &   14.02&           &         &\\
   52726&   2.8481&     0.0769&     0.2880&    13.35&     	&       &\\
   55613&   2.9143&     0.0250&     0.1397&    14.34&      	&       &\\    
   85812&   2.8444&     0.1769&     0.0383&    15.04&      	&       &\\
  141036&   2.9444&     0.0987&     0.0342&    15.78&      	&       &\\   
  160435&   2.8573&     0.0518&     0.0366&    16.35&      	&       & Koronis\\
   
\hline
 \textbf{(221) Eos}\\
\hline
    7472&   3.0102&     0.1389&     0.1563&    12.08&    10.0080&   0.2795&\\
   14390&   3.2449&     0.0844&     0.1165&    12.16&    10.7670&   0.2202&\\
   41243&   2.9893&     0.0766&     0.1906&    14.36&     4.2160&   0.1727& Eos\\
   58860&   3.0561&     0.0704&     0.1598&    14.28&     3.7390&   0.1665& Eos\\
   61877&   3.0607&     0.0700&     0.0302&    14.75&      	&       &\\   
   63085&   3.1380&     0.0852&     0.2059&    14.16&      	&       &\\
   65256&   3.0880&     0.0449&     0.2083&    14.27&      	&       &\\
   97502&   3.0541&     0.2407&     0.1289&    14.43&     3.0460& 0.2751&\\ 
  105041&   3.0403&     0.1583&     0.1847&    14.09&      	&       &\\ 
  106205&   3.0776&     0.1163&     0.0381&    15.50&      	&       &\\  
  117043&   3.0260&     0.0854&     0.0448&    15.17&      	&       &\\
  144548&   3.1207&     0.0419&     0.1840&    15.24&      	&       &\\
  153284&   3.0445&     0.1342&     0.1008&    14.83&     	&       &\\
  177904&   3.1586&     0.1539&     0.0731&    15.30&     5.762 & 0.0403&\\
  189899&   3.0099&     0.1291&     0.2020&    15.20&      	&       &\\
  195852&   3.1326&     0.0525&     0.1546&    15.62&      	&       &\\
  225034&   3.0288&     0.0825&     0.2155&    14.81&      	&       &\\
  261676&   3.0497&     0.1812&     0.0740&    15.67&      	&       &\\
  301865&   3.0634&     0.0864&     0.1979&    16.52&      	&       & Eos\\
 
\hline
 \textbf{(1459) Magnya}\\
\hline  
    1459&   3.1503&     0.2129&     0.2681&    10.24&     29.900&   0.2168&\\
   11465&   3.1026&     0.0149&     0.3480&    12.87&     12.556&   0.0775&\\
   34698&   3.1800&     0.0692&     0.3863&    12.42&      8.070&   0.3920&\\
   63075&   3.1058&     0.0762&     0.2482&    13.97&      3.778&   0.2358&\\
   91159&   3.1946&     0.2738&     0.2773&    14.28&           &         &\\
   92182&   3.1813&     0.1699&     0.2660&    13.70&           &         &\\
  141050&   3.0954&     0.1220&     0.2613&    14.97&      	&         &\\
  188331&   3.1364&     0.2221&     0.2487&    15.21&      	&         &\\
  208324&   3.1307&     0.2703&     0.2870&    14.91&           &         &\\
  311474&   3.0452&     0.1131&     0.2557&    15.28&      	&         &\\

\bottomrule
\end{tabular}
\end{center}
\end{table*}


\bsp

\label{lastpage}

\end{document}